\documentclass{emulateapj}

\usepackage{graphicx}
\usepackage{epsfig}
\usepackage{times}
\usepackage[]{natbib}
\usepackage{url}
\usepackage{color}
\usepackage{amssymb,amsmath}

\newcommand{\kms}{km~s$^{-1}$}
\newcommand{\gcm}{g~cm$^{-3}$}
\newcommand{\ITAacknowledgment}{
This research was supported by the
  Research Council of Norway and by the European Research Council
  under the European Union's Seventh Framework Programme
  (FP7/2007-2013) / ERC Grant agreement nr. 291058. The simulations
  have been run on clusters from the Notur project, and the Pleiades
  cluster through the computing project s1061 from NASA's High End
  Computing Capability (HECC). We thankfully acknowledge the
  computer and supercomputer resources of the Research Council of
  Norway through grant 170935/V30 and through grants of computing time
  from the Programme for Supercomputing, The Swedish 1-m Solar
  Telescope is operated on the island of La Palma by the Institute for
  Solar Physics of Stockholm University in the Spanish Observatorio
  del Roque de los Muchachos of the Instituto de Astrof\'isica de
  Canarias. IRIS is a NASA small explorer mission developed and
  operated by LMSAL with mission operations executed at NASA Ames
  Research center and major contributions to  downlink communications
  funded by ESA and the Norwegian Space Centre.}

\shorttitle{Ellerman bombs and UV bursts}
\shortauthors{Hansteen, Archontis, Pereira, et al.}

\begin{document}

\title{Bombs and flares at the surface and lower atmosphere of the Sun}

%% The command below calls the preprint style
%% which will produce a one-column, single-spaced document.
%% Examples of commands for other substyles follow. Use
%% whichever is most appropriate for your purposes.
%%
%\documentclass[onecolumn]{emulateapj}
%\documentclass[12pt,preprint]{aastex}
%% manuscript produces a one-column, double-spaced document:
%\documentclass[manuscript]{emulateapj}
%% preprint2 produces a double-column, single-spaced document:
%\documentclass[preprint2]{aastex}


\author{V.H. Hansteen\altaffilmark{1}}
\author{V. Archontis\altaffilmark{2}}
\author{T.M.D. Pereira\altaffilmark{1}}
\author{M. Carlsson\altaffilmark{1}}
\author{L. Rouppe van der Voort\altaffilmark{1}}
\author{J. Leenaarts\altaffilmark{3}}


\altaffiltext{1}{Institute of Theoretical Astrophysics, University of
  Oslo, Norway, PB 1029 Blindern, 0315 Oslo, Norway}
\altaffiltext{2}{School of Mathematics and Statistics, St. Andrews University, St. Andrews, KY169SS, UK}
\altaffiltext{3}{Institute for Solar Physics, Dept. of Astronomy,
  Stockholm University, Roslagstullbacken 21 SE-10691 Stockholm, Sweden}


\begin{abstract}
%max 250 words
A spectacular manifestation of solar activity is the appearance of
transient brightenings in the far wings of the H$\alpha$ line, known as
Ellerman bombs (EBs). Recent observations obtained by the Interface
Region Imaging Spectrograph (IRIS) have revealed another type of
plasma ``bombs'' (UV bursts) with high temperatures of perhaps up to 
$8\times 10^4$~K within
the cooler lower solar atmosphere. Realistic numerical modeling
showing such events is needed to explain their nature. Here, we
report on 3D radiative magneto-hydrodynamic simulations of magnetic
flux emergence in the solar atmosphere. We find that ubiquitous
reconnection between emerging bipolar magnetic fields can trigger EBs
in the photosphere, UV bursts in the mid/low chromosphere and small
(nano-/micro-) flares ($10^6$~K) in the upper chromosphere. These results
provide new insights on the emergence and build up of the coronal
magnetic field and the dynamics and heating of the solar surface
and lower atmosphere.

\end{abstract}



\section{Introduction}

The Sun's magnetic field is produced in the solar interior as a result
of dynamo action \citep[][]{1955ApJ...122..293P,1963ARA&A...1...41B}. %Parker 55, Babcock 63
The dynamo-generated field emerges to the
solar surface (photosphere) by magnetic buoyancy and convective
motions. Emergence of magnetic flux is a key process leading to the
formation of active regions, which host the most intense magnetic
activity in the Sun. During  flux emergence, magnetic fields rise into
the photosphere in the form of magnetic bipoles, which can evolve
dynamically to form multi-scale coronal loops over active regions, but
before doing so they must rid themselves of considerable amounts of
mass carried up from below. %Cheung et al. 2010, Rempel & Cheung 2014
%\citep[][]{2010ApJ...720..233C,2014ApJ...785...90R}.

Ellerman bombs \citep[EBs,][]{1917ApJ....46..298E} at the photosphere
and hot explosions ($\sim 8\times 10^4$~K) in the cool ($10^4$~K)
solar atmosphere discovered by IRIS
\cite[UV bursts,][]{2014Sci...346C.315P}, %Peter 2014
are two phenomena typically observed in regions of opposite polarity magnetic
fields, most often during flux emergence
%Georgoulis 2002, Watanabe 2011, Vissers 2013, Rutten 2013
\citep{2002ApJ...575..506G,2011ApJ...736...71W,2013ApJ...774...32V,2013JPhCS.440a2007R,2016ApJ...823..110R},
but also in quiet Sun \citep[][]{2016A&A...592A.100R}.
Several studies
%Isobe 2007, Archontis 2009, Georghoulis 2002, Pariat 2004, Peter
%2014, Watabe 2013, Matsumoto 2008
\citep[][]{2007ApJ...657L..53I,
2009A&A...508.1469A,2002ApJ...575..506G,
2004ApJ...614.1099P,2014Sci...346C.315P,
2011ApJ...736...71W,2008PASJ...60..577M}
have supported the idea that reconnection
between the opposite polarity fields of interacting bipoles may
provide the necessary energy (through conversion from magnetic to
kinetic and thermal energy) to accelerate and heat plasma,
powering these explosive events and triggering small flares $O(10^6~\mathrm{K})$
in the solar atmosphere. EBs and UV bursts share many characteristics
and both occur in emerging active regions, hence their relation to
each other is currently under debate
\citep{2015ApJ...812...11V,2016ApJ...824...96T,2017arXiv170102112D}.

In order to produce the particular characteristics of EB emission
  in the line profile of H$\alpha$, as well as other lines such as the
  \ion{Ca}{2} IR and \ion{Ca}{2} H \& K lines, semi-empirical NLTE
  models have been constructed. In these, the EB atmosphere is modeled
  to consist of a ``hot spot'' of enhanced temperature or density
  spanning a few hundred kilometers near the temperature minimum
  \citep[e.g.][]{2014A&A...567A.110B}, or by using a ``two cloud'' model with an
  absorbing  cool cloud overlying the hot cloud that accounts for the bright
  H$\alpha$ wing emission \citep{2014ApJ...792...13H}. These models
  obtain  reasonable fits to observed
  profiles. \citet{2016A&A...593A..32G} extend this type of analysis
  to the \ion{Mg}{2} h \& k lines observed with IRIS and conclude
  that the bright points observed in these lines could be formed in an
  extended domain spanning the upper photosphere and/or chromosphere
  (400 -- 750~km).

The studies above show that the temperature increases
  responsible for line brightening, both for EBs and phenomena at
  greater heights in the chromosphere such as UV bursts and microflares, 
  presumably are consistent with
  reconnection processes and associated Joule heating. 
  Idealized, but physically
  sophisticated 2D magnetohydrodynamic simulations
  \citep{2011RAA....11..225X,2016ApJ...832..195N}, show that
  reconnection indeed can raise the temperature of the dense plasma to
  $10^4$~K or as high as $\gtrsim 8\times 10^4$~K depending on 
  plasma-$\beta$ (ratio of the gas pressure to the energy density of the magnetic field)
  in the low solar atmosphere 100--600~km above the surface.

In the following, we present a realistic 3D numerical model of
magnetic flux emergence which produces the onset of EBs, UV bursts,
and small chromospheric flares. We study their nature and address the
issue whether EBs and UV bursts are different events.

In our 3D radiative magnetohydrodynamic (MHD) simulations
%Archontis Hansteen 2014, Ortiz et al 2014
\citep{2014ApJ...788L...2A,2014ApJ...781..126O},
a horizontal and uniform magnetic field (flux sheet) emerges from the
convection zone to the photosphere, where it forms a network of
magnetic bipoles. We find that the interaction of bipolar fields, as
driven by resistive emergence into the ambient atmosphere and surface
flows, leads to local plasma heating at various heights. Synthetic
diagnostics from the numerical model reveal that these heating events
match key observed characteristics of EBs, UV bursts and small
chromospheric flares.

\section{The model}
%__________________________________________________________________________________________________
%========================================================
% Figure 1
\begin{figure*}[t]
\centering
\includegraphics[width=0.66\linewidth]{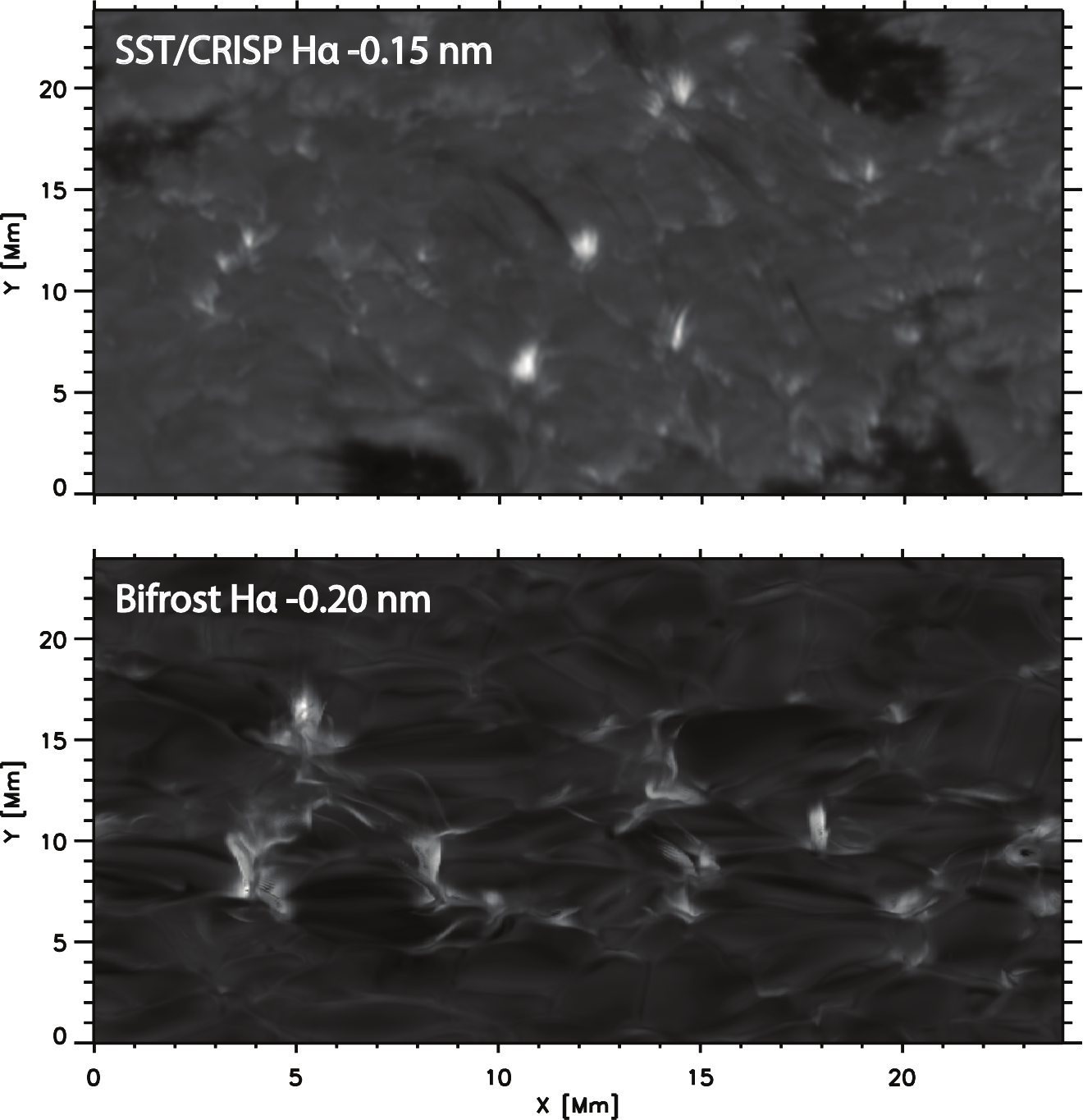}\caption{
Simulated H$\alpha$ line wing emission in EBs closely
resembles flame like structures seen in high resolution ground based
observations. H$\alpha$ intensity in the blue wing (at $-150$~pm)
observed with SST/CRISP (top) and at $-200$~pm in the synthetic
Bifrost images (bottom), at a solar heliocentric angle $\mu=0.5$.
A version of this figure is available as an animation (Movie 1a), showing the
variation of the flame like structures over a period of some 180~s. In
addition an animation (Movie 1b) showing the time variation of the synthetic
H$\alpha$ near the line core (at $+50$~pm) over the same time period is provided. Both animations
are shown at heliocentric angle $\mu=0.5$.}
\label{fig1}
\end{figure*}
%========================================================

%=======================================================
% Figure 2
\begin{figure*}[t]
\centering
\includegraphics[width=0.98\linewidth]{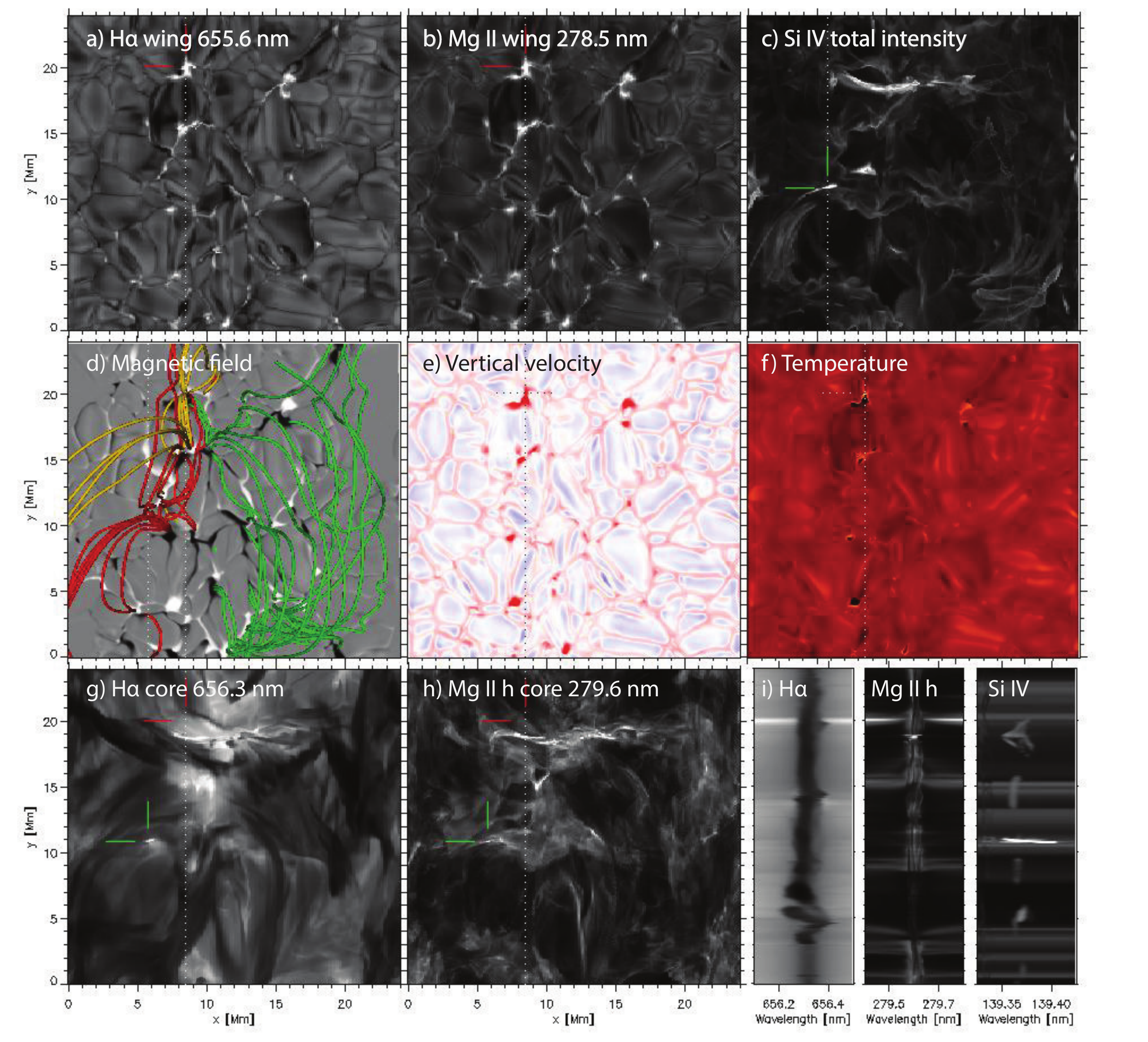}\caption{
Modeled spectra show signatures that resemble the
characteristics of EBs, UV bursts and small flares. H$\alpha$ line
wing (a), \ion{Mg}{2} line wing (b), \ion{Si}{4} total intensity (c); photospheric
magnetogram  ($B_z$) and selected field lines (d), photospheric
velocity $u_z$ (e) and temperature (f); H$\alpha$ line core (g), \ion{Mg}{2}
line core (h), and line profiles of H$\alpha$, \ion{Mg}{2} along
$x=8.5$, and \ion{Si}{4} along $x=6$~Mm (i). The red and green cross
hairs show the location of a simulated EB and UV burst respectively.
In panel d, yellow field lines have been traced
from the EB site (indicated by the red orthogonal lines) and red field
lines from the vicinity of the small flare. The green field lines have
been traced from various magnetic polarities at $z=0$~Mm and for
$x\geq 10$~Mm. An animation (Movie 2) of synthetic H$\alpha$ near the line core
(at $+50$~pm) core seen from 
solar heliocentric angle of $\mu=1$ over a period of 200~s showing formation of opaque dark fibrils over the
emerging flux region is provided.}
\label{fig2}
\end{figure*}
%========================================================

\subsection{Simulations}
The numerical model described in this paper is produced by the Bifrost
code \citep{2011A&A...531A.154G} which solves the
magnetohydrodynamic equations on a 3D Cartesian grid. The modelled
solar atmosphere spans from 2.5~Mm below the solar surface to
14~Mm above the photosphere, and fills 24$\times$24~Mm in the
horizontal direction, i.e. the model covers the upper convection zone
to the lower corona. A grid of $504\times 504\times 496$ was used with a uniform
horizontal resolution of 48~km. The vertical resolution varies from
20~km/cell in the photosphere, chromosphere and transition region to
nearly 100~km/cell in the corona and deeper convective layers, where
the scale height is much larger. Optically thick radiative transfer
and radiative losses from the photosphere and lower chromosphere,
including scattering, are implemented using a short characteristics
scheme \citep[][]{Hayeketal:2010}. Above the lower chromosphere,
radiative losses are parameterized using recipes derived by 1D
non-LTE radiative hydrodynamic simulations
\citep[][]{2012A&A...539A..39C} . Field aligned thermal conduction is
included with the magnetohydrodynamic equations through operator splitting; the
thermal conduction operator is solved using an implicit multi-grid method to allow a
reasonably long time-step.

The model is initialized with a weak uniform oblique magnetic field
($< 0.1$~G) that fills the corona, making an inclination angle of
45 degrees with respect to the $z$-axis. This field is sufficiently
weak that it has no effect on the dynamics of the coronal plasma, but
its non-vertical direction slows the conductive cooling. The coronal
temperature is some 400~kK at the time of flux emergence, overlying a
cool chromosphere mainly heated by acoustic shocks \citep[see][for
further details]{2014ApJ...781..126O}. At the start of the numerical
experiment, a steady state convective equilibrium has been
achieved, the chromosphere is in a quasi steady state, while the
corona is slowly cooling. Then, a non-twisted horizontal flux sheet of strength
$3360$~G is injected at the bottom boundary. The magnetic sheet is
oriented in the $y$ direction filling the area $[x;y]=[3-16;0-24]$~Mm,
for a time period of 1 h 45 m
%Archontis Hansteen 2014, Ortiz et al 2014
\citep[this is the same model as used
in][]{2014ApJ...788L...2A,2014ApJ...781..126O}.

Initially, the
magnetic field emerges to the photosphere, where it stops, since the
field is no longer carried by convective motions nor is buoyant.
As field `piles up' from below, the magnetic pressure increases and
in certain locations plasma-$\beta$ of the
emerging field becomes smaller than one. In these locations magnetic
flux elements can emerge through the photosphere and rise into the outer atmosphere
above. These rising elements expand and rapidly
penetrate and fill the chromosphere
and corona, pushing the pre-existing ambient corona aside.
The heating events that are described
in this paper occur well after (some few thousand seconds) the flux
sheet first encounters the photosphere and are a result of self
interactions between the various magnetic elements that have pierced
the photosphere and are spreading into the chromosphere and corona.

\subsection{Diagnostics}

The response of the solar atmosphere to the emergence and interactions
of the rising magnetic field is studied by computing a number of
spectral lines formed in the photosphere and corona.

Synthetic H$\alpha$ spectra were calculated in 3D non-LTE using the
Multi3d code \citep[][]{2009ASPC..415...87L}. We used a five-level
plus continuum hydrogen model atom and modeled the Ly$\alpha$ and
Ly$\beta$ lines using complete redistribution (CRD) with a Gaussian
profile with Doppler broadening only \citep[][]{2012ApJ...749..136L},
as an approximation to the more time consuming partial redistribution
(PRD) calculations. Radiative transfer calculations were carried out
for a total of 24 angles, using the A4 set
\citep[][]{Carlson1963}. In addition to the snapshot calculated for
display in Figures 1--4, we have calculated H$\alpha$ emission for 
a number of snapshots covering a period of 180~s in order to give an
indication of the time evolution line. For these animations (Movies 1
and 2)  the calculations were run
with reduced horizontal spatial resolution (every second grid point)
to reduce the computational burden. For some snapshots the vertical
grid was interpolated to a finer scale to improve convergence
times. Both of these procedures were found to have a negligible
effect on the line profiles.

The \ion{Mg}{2} synthetic spectra in the vicinity of the h\&k lines
were calculated in 1.5D non-LTE with PRD using the RH 1.5D code
\citep{2001ApJ...557..389U,2015A&A...574A...3P}. In 1.5D each vertical
column from the simulation is treated as an independent 1D
atmosphere. This allows for much faster calculations, at the expense
of less realistic intensities for wavelengths close to the line
cores. This approach is a good approximation for the greater part of
the h\&k line profiles \citep[][]{2013ApJ...772...89L}, making the
computations more tractable than the full 3D PRD calculations
\citep[][]{2016arXiv160605180S}. We used a 10-level plus continuum
\ion{Mg}{2} atom (\citealt{2013ApJ...772...90L}, same setup as
\citealt{2013ApJ...778..143P}, although additional lines of other elements
were not included) and a hybrid angle-dependent PRD \citep[][]{2012A&A...543A.109L}.

The \ion{Si}{4}~139.3744~nm line profiles were also calculated with the RH
1.5D code using a 9 level model atom including 5 levels of \ion{Si}{3},
three levels of \ion{Si}{4} and the ground state of \ion{Si}{5}. The overlapping
\ion{Ni}{2} line was included simultaneously using a four level model atom
including the ground term of \ion{Ni}{2} as two levels, the upper level of
the \ion{Ni}{2}~139.3324~nm line (93 \kms on the blue side of the \ion{Si}{4} line
core) and the ground term of \ion{Ni}{3}. The background continuum of \ion{Si}{1}
was also included in non-LTE employing a 16 level model atom with 15
levels of \ion{Si}{1} and the ground state of \ion{Si}{2}.

\subsection{Observations}

In order to compare and contrast our synthetic spectra with equivalent
observational spectra we show observations obtained at the Swedish 1-meter Solar
Telescope
%Scharmer 2003
\citep[][]{2003SPIE.4853..341S}
using the CRISP instrument
%Scharmer 2008
\citep[][]{2008ApJ...689L..69S}
on September 27 2015 at
10:00:39 UT, centered on the emerging active region
NOAA~12423.
For exploration of the multi-dimensional simulation and observational
data cubes, we made much use of CRISPEX \citep[][]{2012ApJ...750...22V}.

CRISP was running a 3-line program at a temporal cadence of 32~s
during which
the H$\alpha$ spectral line was sampled at 15 line positions between
$\pm 150$~pm. The CRISP program further includes
spectral sampling of the \ion{Fe}{1} 617.3~nm and \ion{Ca}{2} 854.2~nm
lines. CRISP data were processed following the CRISPRED data reduction pipeline
%de al Cruz 2015
\citep[][]{2015A&A...573A..40D},
which includes Multi-Object Multi-Frame Blind Deconvolution image
restoration
%van Noort 2005
\citep[][]{2005SoPh..228..191V}.
The spatial sampling is 0.057~arcsec per pixel, the
diffraction limit at 656.3~nm is 0.14~arcsec. The complete time series
has a duration of 02:43 hours, from 07:47 to 10:30 UT.

Simultaneous data was obtained with NASA's Interface Region Imaging
Spectrograph \citep[IRIS,][]{2014SoPh..289.2733D}  which was observing the
same active region running an observing program
labeled OBS-ID~3620106168. This program comprises a medium-dense
128-step raster with continuous 0.33~arcsec
steps of the 60~arcsec spectrograph slit covering a spatial area of
42.2~arcsec$\times$ 60~arcsec at a temporal
cadence of 10:53~min. The exposure time was 4~s. To increase
signal-to-noise the IRIS spatial sampling was summed on board
($2\times 2$ pixel binning) so that the original IRIS resolution of
0.166~arcsec spatial sampling was reduced to 0.33~arcsec spatial
sampling while the spectral sampling was reduced from $3$~\kms
to $6$~\kms sampling. The IRIS data were calibrated
to ``level 2'', {\it i.e.}, including dark current, flat-field and
geometric correction
%De Pontieu et al
\citep[][]{2014SoPh..289.2733D}. The slit-jaw images (SJI) were
corrected for dark-current and flat-field, as well as internal co-alignment drifts.

Alignment between the SST and IRIS data was done through
cross-correlation of CRISP Ca II 854.2~nm wing images and IRIS
279.6~nm slit jaw images.

\section{Results}

%=======================================================
% Figure Synthetic EB
\begin{figure*}[t]
\centering
\includegraphics[width=0.75\linewidth]{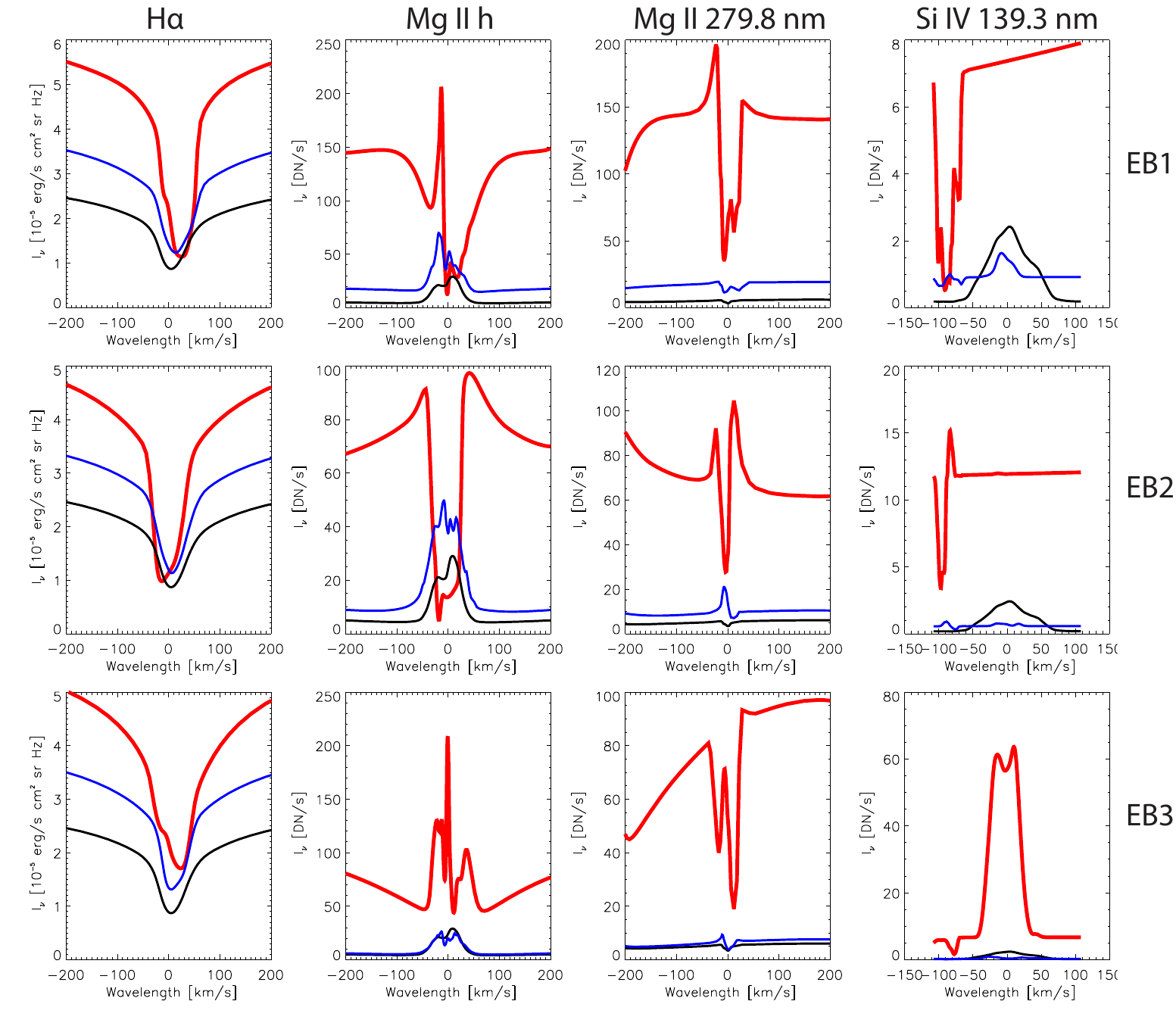}\caption{
Synthetic line profiles from 3 selected Ellerman bombs, as seen from
straight above, showing H$\alpha$ (first column), the \ion{Mg}{2} h
line, \ion{Mg}{2} triplet lines (middle columns) and \ion{Si}{4}
139.3744~nm (fourth column) including \ion{Ni}{2} 139.3324~nm line
located 93~\kms blue-wards of the \ion{Si}{4} line center. Each row
represents a different event. The
red line shows the EB spectra, the black lines show the average
profiles of the entire simulation snapshot, while the blue lines
show average spectra in the vicinity of the EB.}
\label{fig:spectra_EB}
\end{figure*}
%========================================================

%=======================================================
% Figure Synthetic UV
\begin{figure*}[t]
\centering
\includegraphics[width=0.75\linewidth]{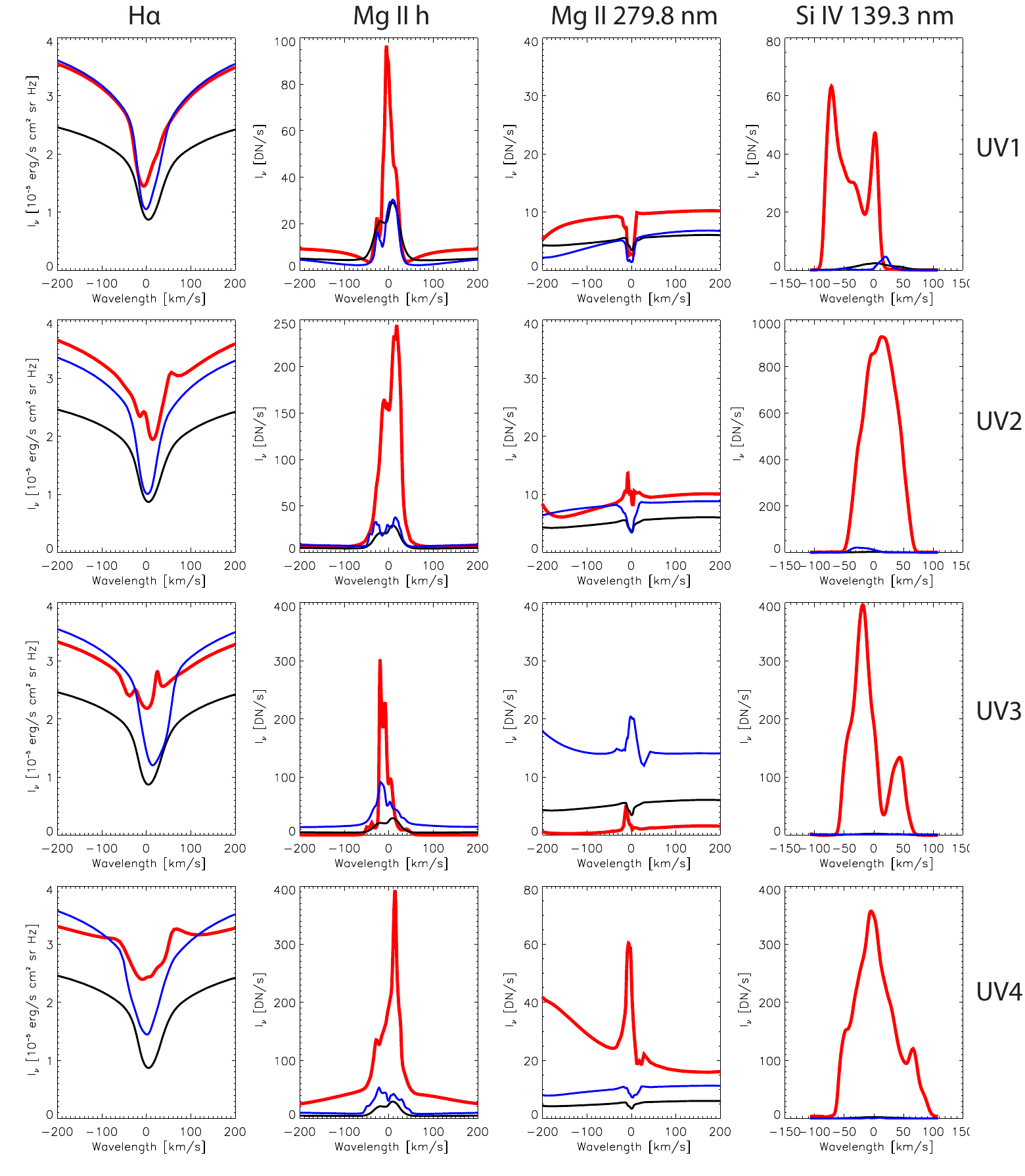}\caption{
Synthetic line profiles from 4 selected UV bursts, as seen from
straight above, showing H$\alpha$ (first column), the \ion{Mg}{2} h
line, \ion{Mg}{2} triplet lines (middle columns) and \ion{Si}{4}
139.3744~nm (fourth column) including \ion{Ni}{2} 139.3324~nm line
located 93~\kms blue-wards of the \ion{Si}{4} line center.  Each row
represents a different event. The 
red line shows the UV burst spectra, the black lines show the average
profiles of the entire simulation snap shot, while the  blue lines
show average spectra in the vicinity of the UV burst. }
\label{fig:spectra_UV}
\end{figure*}
%========================================================

%=======================================================
% Figure Observed Spectra
\begin{figure*}[t]
\centering
\includegraphics[width=0.75\linewidth]{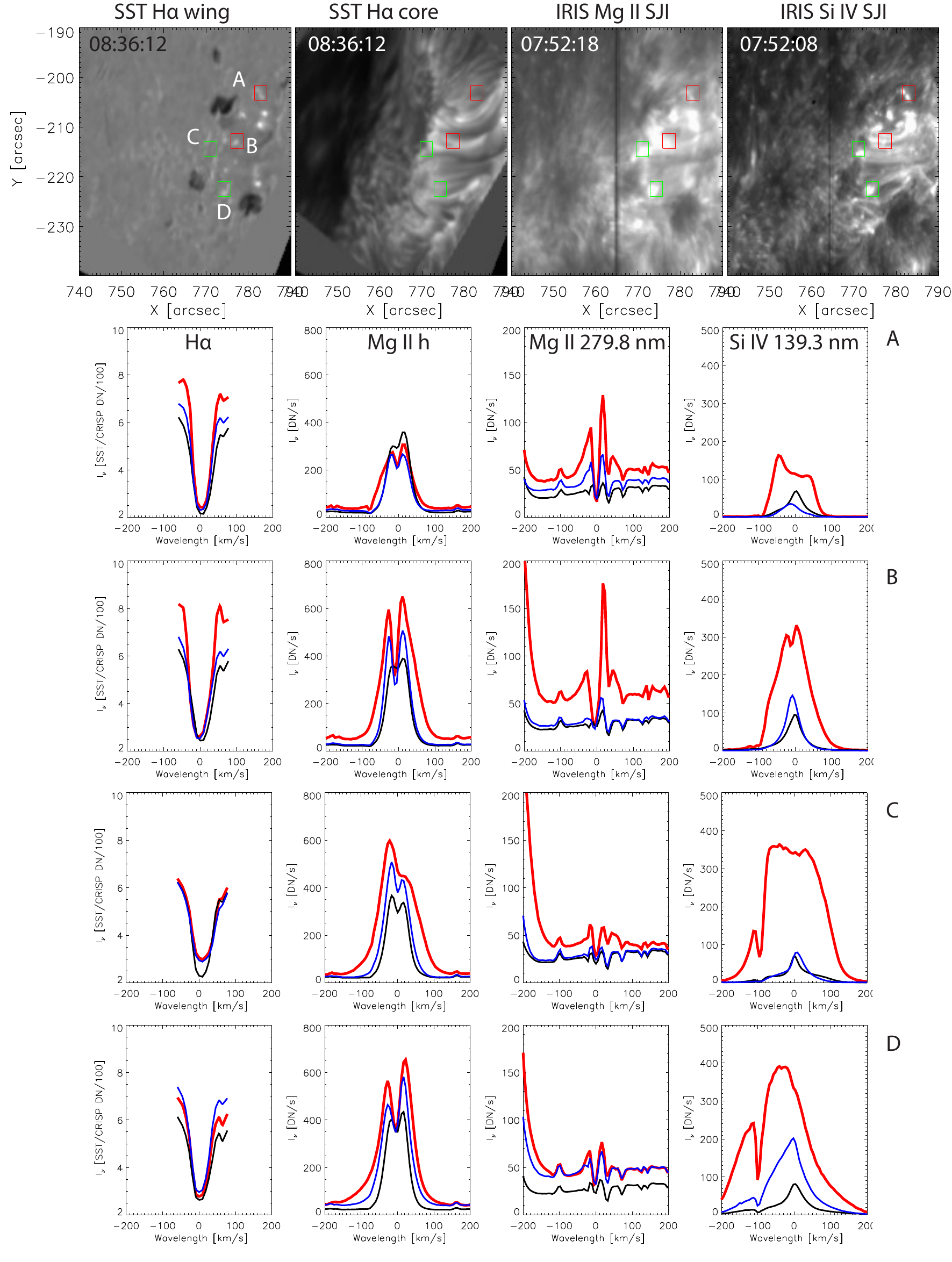}\caption{
SST and IRIS observations of 2 Ellerman bombs and 2 UV bursts seen at at a
solar heliocentric angle $\mu=0.5$. The top panels show the H$\alpha$
line wing and core respectively (left images), as well as the
\ion{Mg}{2} k line and \ion{Si}{4} 1400 slit jaw images (right
column).  The lower panel rows show selected Ellerman bomb and UV
burst, from locations A-D indicated in the top row,
showing H$\alpha$  (first column) the \ion{Mg}{2} h
and \ion{Mg}{2} triplet lines (middle columns) and \ion{Si}{4}
139.3744 nm,  including \ion{Ni}{2} 139.3324 line located 93~\kms
blue-wards of the \ion{Si}{4} line center (right columns). The spectra
A and B are Ellerman bombs, C and D UV bursts. The
red line shows the EB or UV burst spectra, the black lines show the
average profiles of the region surrounding the UV burst or EB, while
the blue lines show time averaged spectra in the vicinity of
the EB or UV burst.}
\label{fig:spectra_obs}
\end{figure*}
%========================================================

A defining observed characteristic of EBs
\citep[][]{2011ApJ...736...71W} is the appearance of
rapidly varying extended flames, or jets, jutting out from the
photospheric network in the H$\alpha$ line, most easily seen when looking off
disk center (see Figure~\ref{fig1}a). These synthetic observations can
also be compared with observations obtained at the Swedish 1-meter Solar
Telescope (Figure~\ref{fig1}b). It is likely that these flames exist below a
`canopy' of dense chromospheric plasma, which prevents any hot
emission from the photosphere to be visible in the H$\alpha$ line
core, formed some $2000$ km above the photosphere
\citep[][]{2012ApJ...749..136L}. Thus, EBs appear far
from line center on both sides of H$\alpha$ line center. \citet{1917ApJ....46..298E}
reported that the enhanced brightness extends at least 0.4-0.5 nm, and
in certain cases attains a width of 3 nm. In the simulations
(Figure~\ref{fig1}a), synthetic images made in the far blue and red wings of H$\alpha$
show several EBs that are rooted in the photospheric network, located
in intergranular lanes between opposite magnetic polarities. They all
possess bright strands of dense plasma, which are vertically oriented,
and they impart an overall flame-like structure to the EBs. These
results come into close agreement with the location, morphology and
dynamics of observed EBs (see also Movie 1a and 1b which show the time
evolution in the line wing and at $+50$~pm from the line core).

%========================================================

Synthetic spectroheliograms from the model (Figure~\ref{fig2}a,~\ref{fig2}b) show that EBs appear as
small (size $1-1.5$~Mm) bright structures in the far wings of H$\alpha$ and
the \ion{Mg}{2} h\&k lines. They occur at the interface between opposite
polarity fields of adjacent emerging bipoles  (Figure~\ref{fig2}d).  Profound
downflows (Figure~\ref{fig2}e) occur on either side of the interface, due to
plasma draining along the flanks of the emerging bipoles. The plasma
at photospheric heights in the interface is heated to $8-9\times 10^3$~K
(Figure~\ref{fig2}f), giving onset to EBs.

The synthetic EBs are not visible at the line core of the H$\alpha$
and \ion{Mg}{2} h lines (Figure~\ref{fig2}g,~\ref{fig2}h) and they are barely visible in the \ion{Si}{4}~139.3~nm
band (Figure~\ref{fig2}c). Just as the H$\alpha$ core, the line core of the
\ion{Mg}{2} h\&k
lines are also mainly formed in the high chromosphere
\citep[][]{2013ApJ...772...89L}. Emission in the \ion{Si}{4}~139.3~nm
line requires temperatures of at least $2\times 10^4$~K in the
dense photosphere or $6-8\times 10^4$~K in the less dense corona
where the coronal approximation is valid. Thus, they
are not responsive to the EBs generated here, which remain
below $< 10^4$~K at photospheric heights. In the far wings of \ion{Mg}{2}
lines, and at the continuum near the \ion{Si}{4} line, (Figure~\ref{fig2}i) lessening
opacity allows brightening from the EBs to become visible through the
overlying canopy, forming a bright `moustache' (see also Figure~\ref{fig:spectra_EB}).

On the other hand, the
H$\alpha$ line core (Figure~\ref{fig2}g) reveals the existence of extended `dark' loops
(e.g. at $x=14$~Mm, $y=0-14$~Mm, see also Movie 2 which shows the time
evolution in H$\alpha$ at $+50$~pm as seen from above), which are lying along the
overarching field-lines (e.g. green field-lines in Figure~\ref{fig2}d) that connect
distant opposite polarities in the emerging flux region. These newly
formed loops carry cool and dense plasma from the
photosphere/chromosphere and they may account for H$\alpha$ arch-filament
systems.

At other locations, there is strong emission at \ion{Si}{4}, with no obvious
corresponding signal in H$\alpha$ (core or wing) and only weak enhancement in
\ion{Mg}{2}. For instance, the \ion{Si}{4} signal for the brightening at
$[x,y]=[8.5,12]$~Mm (Figure~\ref{fig2}c,~\ref{fig2}i) is 200 times stronger than the
average while showing no obvious brightening in H$\alpha$ nor the
\ion{Mg}{2} h line. The absolute intensity of this brightening is of
the same order of magnitude as hot explosions discovered by IRIS
\citep[{\it e.g.} ][see also
Figure \ref{fig:spectra_obs}]{2014Sci...346C.315P}. The \ion{Si}{4} line
profile is also very broad in this location, as discussed below, and
thus, these events may correspond to the observed UV bursts.

\begin{figure*}[t]
\centering
\includegraphics[width=0.5\linewidth]{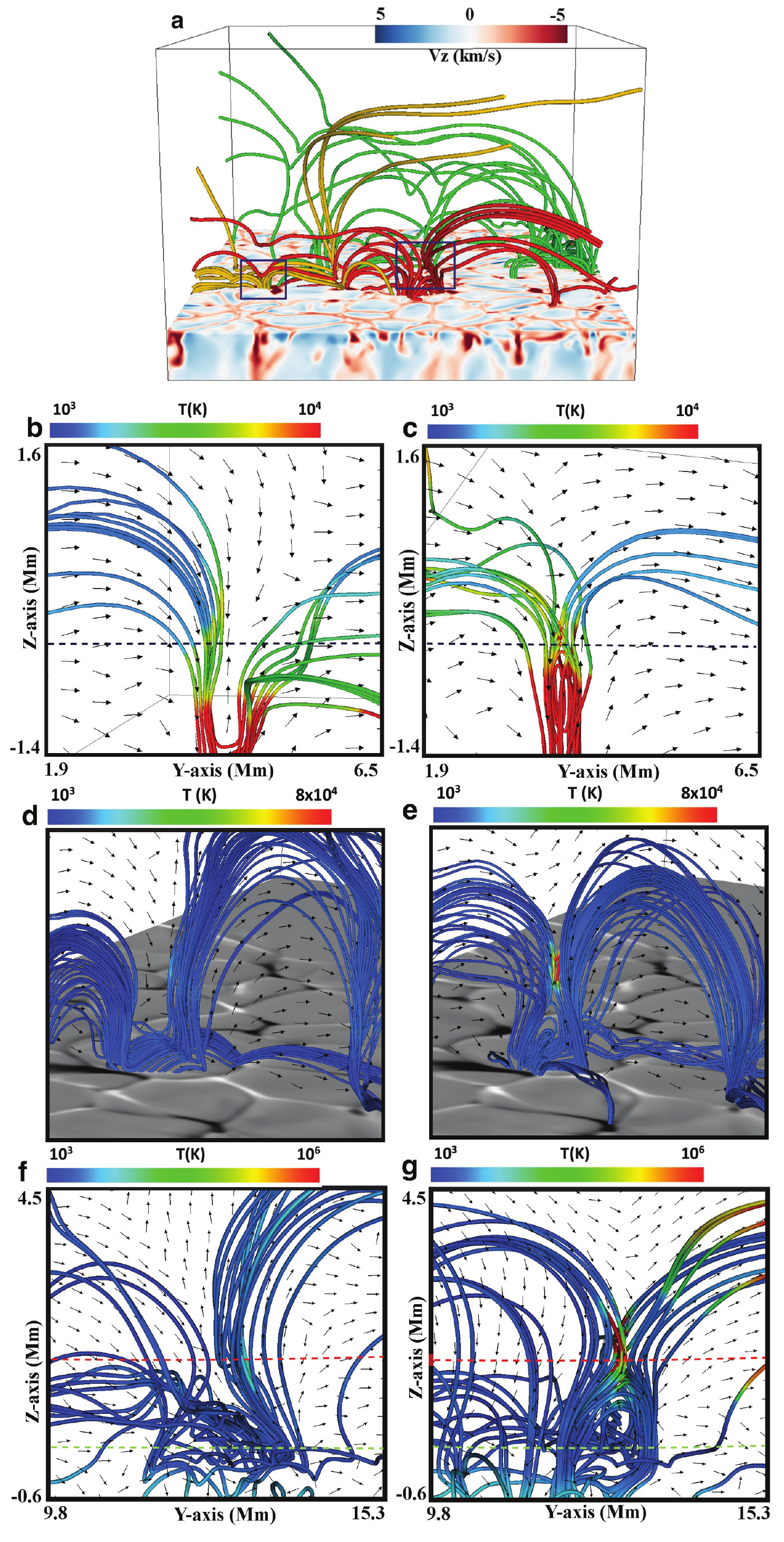}\caption{
Overall field-line topology and the onset of an EB, UV burst
and a small chromospheric flare. a, Selected field-lines showing the
serpentine field at $t=9160$~s. Yellow and red field-lines have been
traced from the areas highlighted by the black insets (left and right
respectively). The image shows the distribution of vertical velocity at
the photosphere (red color indicates downflowing
material). b-c, Close-up of the field-lines, which have been traced
from within the left inset in (a), at $t=8800$~s (b) and $t=9160$~s
(c). They are colored according to the value of the temperature. The
arrows show the projection of the full magnetic field vector at
$x=8.4$~Mm. The horizontal (dashed) line shows the photosphere. The
localized temperature enhancement, just above $z=0$~Mm and $y\approx 3.5$~Mm,
illustrates the location of the EB. d-e, Emergence of $\Omega$-loops,
out of the serpentine field-lines (d, $t=7870$~s), leads to expansion
and reconnection at the chromosphere (e, $t=7930$~s) and the
triggering of a UV burst. The downward-released reconnected
field-lines, build an arcade anchored in the photosphere. The
chromospheric plasma at and close to the reconnection site (e.g. at
the top of the arcade, $y=4.7$~Mm, $z=1.3$~Mm) is heated to high
temperatures, triggering a UV burst. The horizontal slice shows the
$B_z$ distribution at the photosphere (white (black) shows positive
(negative) $B_z$, in the range $[-10^3,10^3$~G]). The arrows show
the full magnetic field vector at $x\approx 8$~Mm. f-g, Close-up of the
field-lines, which have been traced from within the right inset in (a), %
at $t=8800$~s (f) and $t=9250$~s (g). The arrows show the projection
of the full magnetic field vector at $x\approx 6$~Mm. The two horizontal
(dashed) lines show heights at $z=1.75$~Mm and at $z=0$~Mm. Nearby
emerging loops come into contact and reconnect in a similar manner to
the UV burst case. Energy release occurs at the upper chromosphere
(around $z=1.8$~Mm, $y=13$~Mm), where plasma is less dense, leading to
profound plasma heating ($10^6$ K) and the onset of a small
chromospheric flare.}
\label{fig3}
\end{figure*}

\begin{figure}[t]
\centering
\includegraphics[width=0.98\linewidth]{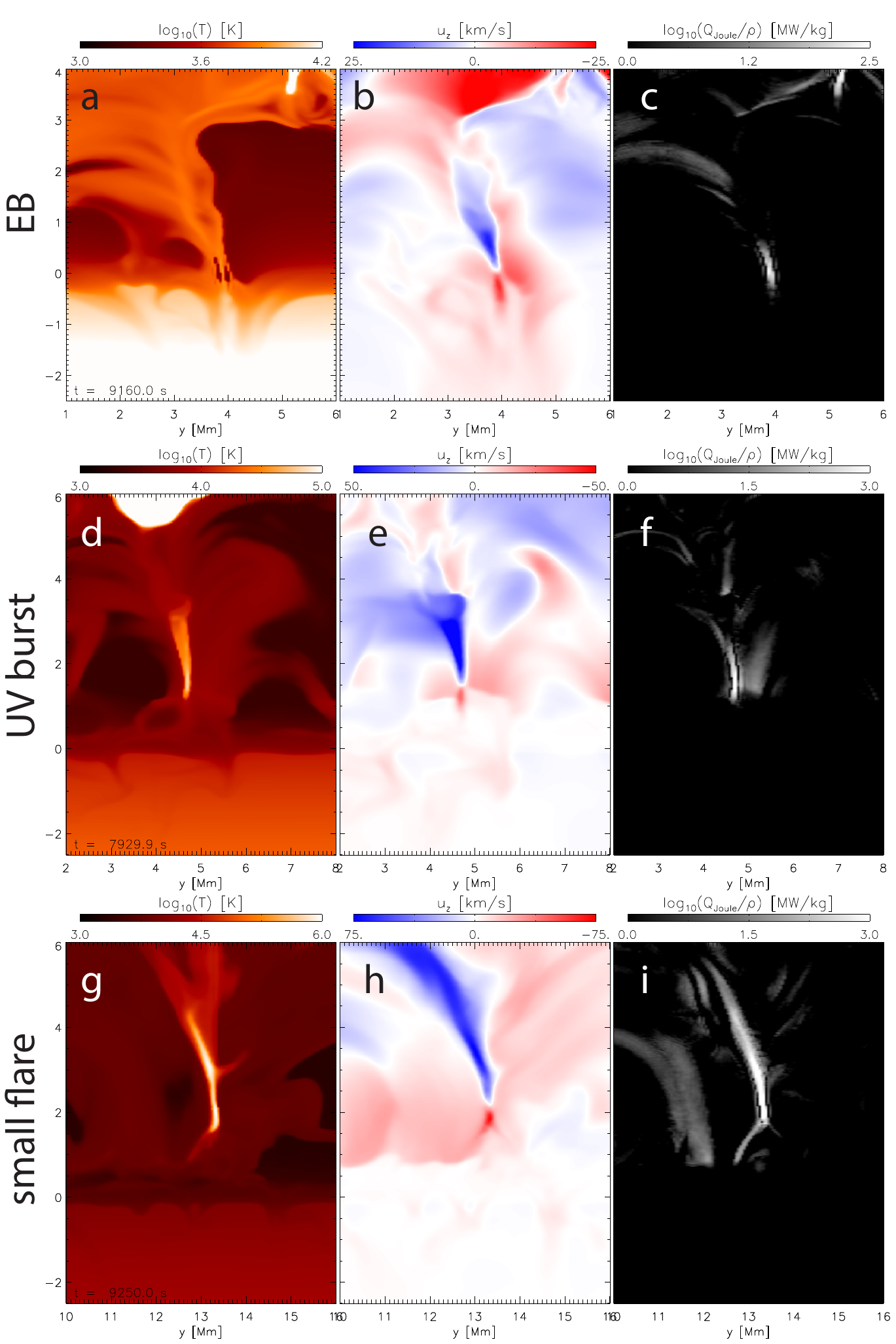}\caption{
Reconnection in the photosphere and chromosphere heats the local
plasma and give rise to fast (supersonic) bi-directional jets at
several heights. 2D cuts at $x=8.43$~Mm, $t=9160$~s (a-c),
$x=8.05$~Mm, $t=7930$~s (d-f) and $x=6.14$~Mm, $t=9250$~s (g-i). Shown
is: temperature (first column), vertical velocity (second column),
Joule dissipation over density (third column). A version of this
figure is available as an animation (Movie 3) showing the time development of
these variable over a period of 500~s.}
\label{fig4}
\end{figure}

Moreover, there are brightenings with very strong emission at \ion{Si}{4}
and considerable emission at the line core of H$\alpha$ and \ion{Mg}{2}  h line
(e.g. at the location indicated by the green orthogonal lines,
Figure~\ref{fig2}c,~\ref{fig2}g,~\ref{fig2}h). This is where plasma at the upper chromosphere is
heated to high temperatures $O(10^6)$~K. These explosions are
associated with the onset of small flares
(Figure~\ref{fig3},~\ref{fig4}).

\subsection{Detailed spectra}

%=======================================================
% Figure 3

The observational signatures of EBs and UV bursts are quite
complicated. To give further insight we here present several synthetic
line spectra from reconnection events occurring in the photosphere and
in the middle to upper chromosphere and compare them with observed
spectra taken at the SST and with IRIS.

In addition to the `flamelike' structure
seen in the wings of H$\alpha$ a number of `secondary' EB
characteristics have been reported \citep{2015ApJ...812...11V}.
In Figure~\ref{fig:spectra_EB}, we present the spectra of several simulated EBs in
H$\alpha$, the \ion{Mg}{2}~h line near 279.6 nm, the
\ion{Mg}{2}~triplet lines near 279.8~nm and the \ion{Si}{4}
139.3744~nm line (including the \ion{Ni}{2} blend at 139.3324~nm). In
Figure~\ref{fig:spectra_UV} the spectra of synthetic UV-bursts are
shown.

Figure~\ref{fig:spectra_EB} shows that the H$\alpha$ line core is
unaffected by the presence of the EB, mainly due to the existence of
an overlying chromospheric canopy above the EB. For all cases, we find
that the line wings are significantly enhanced; by roughly a factor
two when seen from directly above, and that this enhancement extends
sοme 0.1 nm or more on either side of the line core.
As for H$\alpha$, the \ion{Mg}{2} line core is
opaque and emission in the core largely stems from the overlying
(usually cooler) chromosphere, but the overlying material becomes
transparent as one moves away from the central core and we see
brightenings, sometimes asymmetric, in the outer core and wings of the
\ion{Mg}{2}~h line, as well as large brightening in the continuum
surrounding the line.

In most of the simulated EBs, we also find asymmetric emission in the
\ion{Mg}{2}~triplet lines near 279.8~nm.
Emission in these lines is observationally rare and indicates that the chromospheric temperature
is high at low chromospheric heights \citep{2015ApJ...806...14P}.

In general, we do not find any significant enhancement in the
\ion{Si}{4} 139.3744~nm line above our simulated EBs. In the cases
that we have studied, we usually find enhanced continuum emission and
absorption in the \ion{Ni}{2} line (which sometimes shows a complicated line
structure indicative of large relative flows in the chromosphere). In
some instances we find enhanced \ion{Si}{4} emission in the same locations as the
EB, but these are often dominated by heating events, unrelated to the
EB, at greater heights in the chromosphere. Though the emission is
unrelated, the magnetic field structure of these brightenings is
connected, forming active sites in the larger magnetic flux
structure that is emerging through the photosphere. 
This possible connection between disparate emitting sites 
has also been noted observationally \citep{2000ApJ...544L.157Q}. Note that when seen from
the side we occasionally see that the upper part of the EB does
produce emission in \ion{Si}{4}, but in every instance we have found this is
dominated by emission from greater heights when seen from above.

Simulated line profiles from reconnection events that occur higher in
the atmosphere are shown in Figure~\ref{fig:spectra_UV}. The
H$\alpha$ core emission from these events depends on event height and
strength, while the wing emission is always unresponsive. The line
core can either be unresponsive, or in some cases show significant
brightening and complex structure for the strongest most violent
events --- perhaps forming a counterpart to small flares.

The \ion{Mg}{2}~h line core emission is enhanced by a large factor and the
core line profile shows quite complicated structure reflecting the
high velocity jets that form the core of these reconnection events.
As for H${\alpha}$, the \ion{Mg}{2} triplet line emission does not
always respond to these UV-bursts, but occasionally show asymmetric
emission profiles and  emission.

On the other hand, the \ion{Si}{4} emission is greatly enhanced in
some cases. It can increase as much as three orders of magnitude over
the average value and it reflects intensities comparable to those
measured with IRIS. The line profile of the \ion{Si}{4} lines is quite complex,
but in general shows bi-directional Doppler shifts with amplitudes up
to 100~\kms. The \ion{Si}{4} line profile is very broad, of
order 100 \kms, indicating highly supersonic velocities for the lines
formed at the greatest heights (the speed of
sound is of order 30~\kms\ at the temperature of \ion{Si}{4}
emission). The cases of high \ion{Si}{4} intensities we
report here are all formed in the middle to upper chromosphere and we
find that the line has moderate opacity, with $\tau\sim 10$. We would
expect \ion{Si}{4} to be even thicker, were the line formed deeper in the atmosphere, {\it e.g.} in
connection with an EB.

%=======================================================
% Figure 4
%========================================================

To make a comparison between the simulated spectra and the
observations, we present line profiles of a number of EBs and UV
bursts observed with the SST and IRIS on September 27 2015 starting
at 07:52:39 UT in the vicinity of active region NOAA~12423.
Two sets of two spectra are presented in Figure~\ref{fig:spectra_obs}.

In the first two cases, (A) and (B), two Ellerman
bombs were selected by inspecting the H$\alpha$ wing spectroheliogram
for bright flamelike structures and then plotting the spectra of
H$\alpha$ and the concurrent IRIS UV lines in the same region. In
addition, to give an idea of the temporal variability of these lines
in that region, we plot the average profile and spectra at a temporal
cadence of some 22~minutes for the entire observational run lasting some
4.5~hours. H$\alpha$ responds as expected, with bright wings and unresponsive
core emission. The \ion{Mg}{2} h line core emission is unresponsive,
while the h2 peaks are broadened, sometimes (case B), showing
increased emission, as is also the case in the continuum where the
intensity increases by almost a factor 2. The
\ion{Mg}{2} triplet lines show both enhanced emission and asymmetric
profiles. In the \ion{Si}{4} line we find broad $O(200\mathrm{km~s}^{-1})$
profiles and enhanced emission, but less than in the UV burst cases
discussed below.

Likewise, cases (C) and (D) were selected by searching for bright
\ion{Si}{4} SJI emission and then plotting H$\alpha$ and the other
IRIS lines concurrently, including average and time variability spectra
for reference. The \ion{Si}{4} line spectra are extremely broad
$O(300\mathrm{km~s}^{-1})$, sometimes asymmetric (case D), and \ion{Ni}{2}
absorption is strong in both examples. H$\alpha$ emission, on the
other hand, seems quite unresponsive. We note that there is a
Ellerman bomb-like brightening that occurs in case D, but it appears
some 66~minutes after the original \ion{Si}{4} brightening and is
slightly offset from the \ion{Si}{4} SJ emission, thus likely not
representing  the same event. \ion{Mg}{2} h line emission is enhanced, both in the h2 peaks
and (case C) in the h3 line core, as well as being asymmetric. The
continuum emission is slightly enhanced. The
\ion{Mg}{2} triplet lines are in emission, though to a much smaller
extent than as seen in the EB cases (A and B).

The time cadence of this observation is unfortunately too low to show
the time evolution of these events clearly, obtaining such data should
be a goal for future studies.

\subsection{Dynamics and energetics}

To understand the mechanism for the onset of the EBs, UV bursts, and
small flares in our experiment, we study the topology of the magnetic
field (Figure~\ref{fig3}a). Initially, the sub-photospheric magnetic field is
pulled up by convective up-flows and buoyancy (where the field is
strong) and it is dragged down by downdrafts, developing an undulating
shape. When the strongest field breaks through the surface, small
($1-2$~Mm) magnetic bipoles appear at the photosphere. Eventually, the field
expands into the ambient atmosphere forming $\Omega$-like loops, which may
reconnect with each other, creating longer magnetic structures (i.e. the
arch-filaments shown in Figure~\ref{fig2}). As a result, the emerging magnetic
field develops an overall ``sea-serpent'' configuration of loops with
increasingly larger scale with height.

The EBs are formed at the photosphere, between the opposite polarity
field-lines of adjacent emerging loops (Figure~\ref{fig3}b,~\ref{fig3}c). As the U-like
parts of the undulating field are pulled down by convective
down-flows, the nearby legs of the adjacent loops are pressed together
and their field-lines reconnect, which leads to local plasma heating
and the onset of EBs. A side effect of this process is the unloading
of mass from the serpentine field
%Isobe 2007, Archontis 2009, Cheung 2010, Tortosa-Andreu 2009
\citep[][]{2007ApJ...657L..53I,2009A&A...508.1469A,2010ApJ...720..233C,2009A&A...507..949T}.
Reconnection above the density-loaded U-dips, forms twisted magnetic structures
(O-shaped, as projected onto the $yz$-plane), which are dragged into the
convection zone, together with their attached heavy plasma. Some
amount of cool plasma is carried upward with the expanding field
forming the opaque chromospheric canopy.

A similar mechanism occurs at larger heights, powering UV bursts (at
$z\sim 1.5$~Mm, Figure~\ref{fig3}d, \ref{fig3}e) and small flares (at $z\sim 1.8$~Mm,
Figure~\ref{fig3}f,~\ref{fig3}g). The lateral expansion of adjacent $\Omega$-loops, brings their
oppositely-directed stressed fields into contact, leading to
reconnection at their interface and heating of chromospheric plasma to
high temperatures ($7-8\times 10^4$~K for the UV bursts and $\sim
1$~MK for the small flares).

Figure~\ref{fig4} shows vertical cuts at the locations of an EB (a-c), UV burst
(d-f) and small chromospheric flare (g-i). In all cases, strong
currents are built up at the interfaces between the interacting
magnetic fields. Reconnection at the current layers leads to plasma
heating via Joule dissipation (c, f, i, Movie 3 which shows the time
evolution of these quantities) and to the emission
of bi-directional flows (b, e, h) with velocities comparable to the
local Alfv{\'e}n speed. Note that the velocities are asymmetric, with
the upward propagating portion having substantially higher velocity
\citep[see {\it e.g.}][]{2016arXiv161001321L}.
The afore mentioned events have lifetimes of at least 120-180~s.

Above the EB, dense ($\sim 10^{-8}$~\gcm) and cool
($\sim 7.5\times 10^3$~K) plasma is
emitted along the reconnection jet with a velocity of $\sim 20$~\kms.
Some $1300$~km higher, at the site of the UV burst, dense
chromospheric material ($\sim 5\times 10^{-13}$~\gcm) is heated to $\sim 7.5\times 10^4$~K
(d). Before the burst, the local (unperturbed) plasma has similar
density but it is 10 times cooler. The total intensity computed in the
\ion{Si}{4} line also shows that the UV burst occurs at chromospheric
heights ($z=1-2$~Mm) in this case. The
reconnection jets run with speeds $40-70$~\kms (e). The upward jet
shoots cool ($1-2\times 10^4$~K) chromospheric plasma to coronal heights
(reaches $z\geq 4$~Mm at 7980~s). The downward jet terminates at the apex of
the post-reconnection arcade (at $z\sim 1$~Mm) where the plasma is heated by
compression. Our results suggest that most UV bursts may
originate from low/mid chromospheric plasma and are not necessarily photospheric as
previously suggested \citep[][]{2014Sci...346C.315P}.

When stronger magnetic fields ($50-100$~G) reconnect in the upper
chromosphere (e.g. $z\sim 2.1$~Mm, panel g), the local plasma ($\sim
1-1.5\times 10^{-13}$~\gcm) is heated to even
higher temperatures ($O(10^6)$~K). The local
plasma-$\beta$, within the volume ($\sim 700 \times 500 \times 1000$~km$^3$) of the profound
heating, is $\sim 0.02$, which explains the increase of temperature by a
factor of $\sim 1/\beta$. The hot and dense reconnection jets are
emitted upward (up to $z\sim 4$~Mm) and downward (to $\sim 1.5$~Mm) with maximum
speeds of some $80$~\kms.

The energetics of our synthetic EBs, UV bursts, and small flares can
be estimated as follows: The (average) magnetic field close to the EB
described in the main text is $\sim 600$~G. The total magnetic energy
stored in the EB's volume ($\sim 500~\mathrm{km} \times 500~\mathrm{km} \times 400~\mathrm{km})$ is $\sim
1.5\times 10^{20}$~J. The total thermal energy content is $\sim
10^{20}$~J. These values lie within the expected energy  range for EBs.
The magnetic field strength near the UV burst is $15-20$~G and
plasma $\beta\approx 0.1$. Thus, the total magnetic energy dumped in
the burst's volume ($\sim 700~\mathrm{km}\times 500~\mathrm{km}\times
800~\mathrm{km}$)  and the involved thermal energy are up to
$3.8\times 10^{17}$~J and $2-3\times 10^{17}$~J respectively, which
are much less than those computed for the synthetic EB.

In the upper chromosphere, the flaring event described in the main
text has a magnetic field  of $50-100$~G. The total stored energy is
$O(10^{18})$~J, which lies within the nanoflare energy
regime. According to the above values, the heated plasma around the
reconnection  site may account for an EUV flare and the emitted jet
for an EUV jet.

\section{Discussion and Conclusions}

The simulations and synthetic diagnostics presented here show that
many, though perhaps not all, observed characteristics of Ellerman
bombs and UV bursts can be reproduced within the confines of a
relatively simple scenario where a untwisted flux sheet is allowed to
emerge through the solar atmosphere and expand into the overlying
atmosphere.

While H$\alpha$ and the \ion{Mg}{2} triplet lines appear very similar
to the observed spectra presented in Fig~\ref{fig:spectra_obs}, the
\ion{Mg}{2} h (and k) lines are not as easy to characterize. 
The observed EB spectra show enhanced h2 and slightly enhanced h3 
peaks and line wings. The synthetic spectra show strongly enhanced wings 
but only sometimes enhanced cores and strong asymmetries. 
We suspect that the difference in viewing angle ($\mu=1$ for the
simulations, $\mu=0.5$ for the observations) 
is the cause of the difference in the wings: \citet{2016A&A...593A..32G} observe 
EBs close to disk center and they observe strongly enhanced wings. 
The observed UV burst spectra show strong enhancement for the h2 cores,
sometimes enhanced h3 and intensity increases in the continuum. This
is not too different from what is found in the synthetic spectra,
though the increase in core width is much greater in many observed
spectra. It should be noted that the variations in \ion{Mg}{2} h\&k line
response to UV bursts is very large as can be ascertained also from other
sources \citep[][]{2016A&A...593A..32G,2015ApJ...812...11V}.

% While H$\alpha$ and the \ion{Mg}{2} triplet lines seem very similar
% to observed spectra, the situation for the \ion{Mg}{2} h (and k) lines
% is much less clear. This may be due to a number of factors, but it is
% worth noting that this line is very difficult to reproduce also for
% quiet Sun and network models \citep{2016A&A...585A...4C}. The spectra
% of EBs and UV bursts are formed from a combination of the undisturbed
% background atmosphere and the event itself, and thus the
% difficulties encountered here could be a reflection of a poor
% chromospheric background  model.

The synthetic \ion{Si}{4} spectra show an intriguing similarity to the
observed line profiles, but there are also differences that could
point the way towards a better understanding of the solar atmosphere
in general and EBs and UV bursts in particular. We found no
\ion{Ni}{2} absorption in those \ion{Si}{4} line profiles that
extend to more than 100~\kms in the blue. This could be due to several
factors: 1) there is not enough cool material above the sites of
strong Si IV emission in the model to give sufficient opacity in the
\ion{Ni}{2} line, 2) the simulated EBs and deep UV bursts do not heat
the denser low atmosphere material to high enough temperatures to
cause \ion{Si}{4} emission and/or 3) the velocities in deeper UV
bursts are not great enough to give significant emission at
a blue-shift of 100~\kms in locations where the opacity of the
overlying material is great enough. We note that reconnection in this
model is mediated by the use of an artificial hyper-diffusive
operator, necessary to prevent the collapse of current sheets to
smaller widths than the grid size (48~km). The effective diffusivity
on the real Sun may allow current sheets to become thinner, with
attendant higher temperatures. Not covered in this study is the
possibility that reconnection, even at low heights in the atmosphere, 
produces non-thermal electrons in
sufficient quantities to perturb emission. \citet{1998A&A...332..761D}
and \citet{2006ApJ...643.1325F} show that this process may be of
interest. In particular these authors find that the the temperature
rise needed to explain EB emission is somewhat lower than that found in
models without particle beams. 
On the other hand it is also likely
that since EBs have a fairly large vertical extent, also with respect
to Joule heating (see Figure~\ref{fig4}), a slightly altered magnetic
topology could lead to \ion{Si}{4} emission from the upper part of EBs
rooted in the photosphere, even in models with relatively poor
resolution, such as this one. The study of EBs and UV bursts, by
comparing observational and  synthetic diagnostics, thus presents a
unique opportunity to study reconnection, and the physical parameters
controlling reconnection, at many heights in the solar
atmosphere.

In order to form the coronal magnetic field, emerging magnetic flux
must break through the photosphere, while ridding itself of
considerable mass. This process needs reconnection in order to
proceed, occurring at several heights as the field rises into the upper
atmosphere forming continually longer loops.
The above results reveal that reconnection between stressed magnetic
fields in an emerging flux region can trigger three types of solar
explosive events (EBs, UV bursts, and small chromospheric flares) of
different origins. We conjecture that a similar mechanism can form a
continuum of explosions across the solar atmosphere, depending on the
local properties of the plasma and the amount of available energy
involved in the process whose visibility is determined by the amount of
overlying opaque (and cool) material.

\acknowledgments
\ITAacknowledgment
\bibliographystyle{apj}
%\bibliographystyle{aa}
%\bibliography{solarrefs}

\begin{thebibliography}{50}
\expandafter\ifx\csname natexlab\endcsname\relax\def\natexlab#1{#1}\fi

\bibitem[{{Archontis} \& {Hansteen}(2014)}]{2014ApJ...788L...2A}
{Archontis}, V. \& {Hansteen}, V. 2014, \apjl, 788, L2

\bibitem[{{Archontis} \& {Hood}(2009)}]{2009A&A...508.1469A}
{Archontis}, V. \& {Hood}, A.~W. 2009, \aap, 508, 1469

\bibitem[{{Babcock}(1963)}]{1963ARA&A...1...41B}
{Babcock}, H.~W. 1963, \araa, 1, 41

\bibitem[{{Berlicki} \& {Heinzel}(2014)}]{2014A&A...567A.110B}
{Berlicki}, A. \& {Heinzel}, P. 2014, \aap, 567, A110

\bibitem[{{Carlson}(1963)}]{Carlson1963}
{Carlson}, B.~G. 1963, Methods in Computational Physics, ed. B.~{Alder},
  S.~{Fernbach}, \& M.~{Rotenberg}, Vol.~1 (New York: Academic)

\bibitem[{{Carlsson} \& {Leenaarts}(2012)}]{2012A&A...539A..39C}
{Carlsson}, M. \& {Leenaarts}, J. 2012, \aap, 539, A39

\bibitem[{{Cheung} {et~al.}(2010){Cheung}, {Rempel}, {Title}, \&
  {Sch{\"u}ssler}}]{2010ApJ...720..233C}
{Cheung}, M.~C.~M., {Rempel}, M., {Title}, A.~M., \& {Sch{\"u}ssler}, M. 2010,
  \apj, 720, 233

\bibitem[{{Danilovic}(2017)}]{2017arXiv170102112D}
{Danilovic}, S. 2017, ArXiv e-prints [\eprint[arXiv]{1701.02112}]

\bibitem[{{de la Cruz Rodr{\'{\i}}guez} {et~al.}(2015){de la Cruz
  Rodr{\'{\i}}guez}, {L{\"o}fdahl}, {S{\"u}tterlin}, {Hillberg}, \& {Rouppe van
  der Voort}}]{2015A&A...573A..40D}
{de la Cruz Rodr{\'{\i}}guez}, J., {L{\"o}fdahl}, M.~G., {S{\"u}tterlin}, P.,
  {Hillberg}, T., \& {Rouppe van der Voort}, L. 2015, \aap, 573, A40

\bibitem[{{De Pontieu} {et~al.}(2014){De Pontieu}, {Title}, {Lemen}, {Kushner},
  {Akin}, {Allard}, {Berger}, {Boerner}, {Cheung}, {Chou}, {Drake}, {Duncan},
  {Freeland}, {Heyman}, {Hoffman}, {Hurlburt}, {Lindgren}, {Mathur}, {Rehse},
  {Sabolish}, {Seguin}, {Schrijver}, {Tarbell}, {W{\"u}lser}, {Wolfson},
  {Yanari}, {Mudge}, {Nguyen-Phuc}, {Timmons}, {van Bezooijen}, {Weingrod},
  {Brookner}, {Butcher}, {Dougherty}, {Eder}, {Knagenhjelm}, {Larsen},
  {Mansir}, {Phan}, {Boyle}, {Cheimets}, {DeLuca}, {Golub}, {Gates}, {Hertz},
  {McKillop}, {Park}, {Perry}, {Podgorski}, {Reeves}, {Saar}, {Testa}, {Tian},
  {Weber}, {Dunn}, {Eccles}, {Jaeggli}, {Kankelborg}, {Mashburn}, {Pust},
  {Springer}, {Carvalho}, {Kleint}, {Marmie}, {Mazmanian}, {Pereira}, {Sawyer},
  {Strong}, {Worden}, {Carlsson}, {Hansteen}, {Leenaarts}, {Wiesmann},
  {Aloise}, {Chu}, {Bush}, {Scherrer}, {Brekke}, {Martinez-Sykora}, {Lites},
  {McIntosh}, {Uitenbroek}, {Okamoto}, {Gummin}, {Auker}, {Jerram}, {Pool}, \&
  {Waltham}}]{2014SoPh..289.2733D}
{De Pontieu}, B., {Title}, A.~M., {Lemen}, J.~R., {et~al.} 2014, \solphys, 289,
  2733

\bibitem[{{Ding} {et~al.}(1998){Ding}, {Henoux}, \&
  {Fang}}]{1998A&A...332..761D}
{Ding}, M.~D., {Henoux}, J.-C., \& {Fang}, C. 1998, \aap, 332, 761

\bibitem[{{Ellerman}(1917)}]{1917ApJ....46..298E}
{Ellerman}, F. 1917, \apj, 46, 298

\bibitem[{{Fang} {et~al.}(2006){Fang}, {Tang}, {Xu}, {Ding}, \&
  {Chen}}]{2006ApJ...643.1325F}
{Fang}, C., {Tang}, Y.~H., {Xu}, Z., {Ding}, M.~D., \& {Chen}, P.~F. 2006,
  \apj, 643, 1325

\bibitem[{{Georgoulis} {et~al.}(2002){Georgoulis}, {Rust}, {Bernasconi}, \&
  {Schmieder}}]{2002ApJ...575..506G}
{Georgoulis}, M.~K., {Rust}, D.~M., {Bernasconi}, P.~N., \& {Schmieder}, B.
  2002, \apj, 575, 506

\bibitem[{{Grubecka} {et~al.}(2016){Grubecka}, {Schmieder}, {Berlicki},
  {Heinzel}, {Dalmasse}, \& {Mein}}]{2016A&A...593A..32G}
{Grubecka}, M., {Schmieder}, B., {Berlicki}, A., {et~al.} 2016, \aap, 593, A32

\bibitem[{{Gudiksen} {et~al.}(2011){Gudiksen}, {Carlsson}, {Hansteen}, \&
  et~al.}]{2011A&A...531A.154G}
{Gudiksen}, B.~V., {Carlsson}, M., {Hansteen}, V.~H., \& et~al. 2011, \aap,
  531, A154

\bibitem[{{Hayek} {et~al.}(2010){Hayek}, {Asplund}, {Carlsson}, {Trampedach},
  {Collet}, {Gudiksen}, {Hansteen}, \& {Leenaarts}}]{Hayeketal:2010}
{Hayek}, W., {Asplund}, M., {Carlsson}, M., {et~al.} 2010, \aap, 517, A49

\bibitem[{{Hong} {et~al.}(2014){Hong}, {Ding}, {Li}, {Fang}, \&
  {Cao}}]{2014ApJ...792...13H}
{Hong}, J., {Ding}, M.~D., {Li}, Y., {Fang}, C., \& {Cao}, W. 2014, \apj, 792,
  13

\bibitem[{{Isobe} {et~al.}(2007){Isobe}, {Tripathi}, \&
  {Archontis}}]{2007ApJ...657L..53I}
{Isobe}, H., {Tripathi}, D., \& {Archontis}, V. 2007, \apjl, 657, L53

\bibitem[{{Leenaarts} \& {Carlsson}(2009)}]{2009ASPC..415...87L}
{Leenaarts}, J. \& {Carlsson}, M. 2009, in Astronomical Society of the Pacific
  Conference Series, Vol. 415, The Second Hinode Science Meeting: Beyond
  Discovery-Toward Understanding, ed. B.~{Lites}, M.~{Cheung}, T.~{Magara},
  J.~{Mariska}, \& K.~{Reeves}, 87

\bibitem[{{Leenaarts} {et~al.}(2012{\natexlab{a}}){Leenaarts}, {Carlsson}, \&
  {Rouppe van der Voort}}]{2012ApJ...749..136L}
{Leenaarts}, J., {Carlsson}, M., \& {Rouppe van der Voort}, L.
  2012{\natexlab{a}}, \apj, 749, 136

\bibitem[{{Leenaarts} {et~al.}(2012{\natexlab{b}}){Leenaarts}, {Pereira}, \&
  {Uitenbroek}}]{2012A&A...543A.109L}
{Leenaarts}, J., {Pereira}, T., \& {Uitenbroek}, H. 2012{\natexlab{b}}, \aap,
  543, A109

\bibitem[{{Leenaarts} {et~al.}(2013{\natexlab{a}}){Leenaarts}, {Pereira},
  {Carlsson}, {Uitenbroek}, \& {De Pontieu}}]{2013ApJ...772...89L}
{Leenaarts}, J., {Pereira}, T.~M.~D., {Carlsson}, M., {Uitenbroek}, H., \& {De
  Pontieu}, B. 2013{\natexlab{a}}, \apj, 772, 89

\bibitem[{{Leenaarts} {et~al.}(2013{\natexlab{b}}){Leenaarts}, {Pereira},
  {Carlsson}, {Uitenbroek}, \& {De Pontieu}}]{2013ApJ...772...90L}
{Leenaarts}, J., {Pereira}, T.~M.~D., {Carlsson}, M., {Uitenbroek}, H., \& {De
  Pontieu}, B. 2013{\natexlab{b}}, \apj, 772, 90

\bibitem[{{Libbrecht} {et~al.}(2016){Libbrecht}, {Joshi}, {de la Cruz
  Rodr{\'{\i}}guez}, {Leenaarts}, \& {Asensio Ramos}}]{2016arXiv161001321L}
{Libbrecht}, T., {Joshi}, J., {de la Cruz Rodr{\'{\i}}guez}, J., {Leenaarts},
  J., \& {Asensio Ramos}, A. 2016, ArXiv e-prints [\eprint[arXiv]{1610.01321}]

\bibitem[{{Matsumoto} {et~al.}(2008){Matsumoto}, {Kitai}, {Shibata}, {Nagata},
  {Otsuji}, {Nakamura}, {Watanabe}, {Tsuneta}, {Suematsu}, {Ichimoto},
  {Shimizu}, {Katsukawa}, {Tarbell}, {Lites}, {Shine}, \&
  {Title}}]{2008PASJ...60..577M}
{Matsumoto}, T., {Kitai}, R., {Shibata}, K., {et~al.} 2008, \pasj, 60, 577

\bibitem[{{Ni} {et~al.}(2016){Ni}, {Lin}, {Roussev}, \&
  {Schmieder}}]{2016ApJ...832..195N}
{Ni}, L., {Lin}, J., {Roussev}, I.~I., \& {Schmieder}, B. 2016, \apj, 832, 195

\bibitem[{{Ortiz} {et~al.}(2014){Ortiz}, {Bellot Rubio}, {Hansteen}, {de la
  Cruz Rodr{\'{\i}}guez}, \& {Rouppe van der Voort}}]{2014ApJ...781..126O}
{Ortiz}, A., {Bellot Rubio}, L.~R., {Hansteen}, V.~H., {de la Cruz
  Rodr{\'{\i}}guez}, J., \& {Rouppe van der Voort}, L. 2014, \apj, 781, 126

\bibitem[{{Pariat} {et~al.}(2004){Pariat}, {Aulanier}, {Schmieder},
  {Georgoulis}, {Rust}, \& {Bernasconi}}]{2004ApJ...614.1099P}
{Pariat}, E., {Aulanier}, G., {Schmieder}, B., {et~al.} 2004, \apj, 614, 1099

\bibitem[{{Parker}(1955)}]{1955ApJ...122..293P}
{Parker}, E.~N. 1955, \apj, 122, 293

\bibitem[{{Pereira} {et~al.}(2015){Pereira}, {Carlsson}, {De Pontieu}, \&
  {Hansteen}}]{2015ApJ...806...14P}
{Pereira}, T.~M.~D., {Carlsson}, M., {De Pontieu}, B., \& {Hansteen}, V. 2015,
  \apj, 806, 14

\bibitem[{{Pereira} {et~al.}(2013){Pereira}, {Leenaarts}, {De Pontieu},
  {Carlsson}, \& {Uitenbroek}}]{2013ApJ...778..143P}
{Pereira}, T.~M.~D., {Leenaarts}, J., {De Pontieu}, B., {Carlsson}, M., \&
  {Uitenbroek}, H. 2013, \apj, 778, 143

\bibitem[{{Pereira} \& {Uitenbroek}(2015)}]{2015A&A...574A...3P}
{Pereira}, T.~M.~D. \& {Uitenbroek}, H. 2015, \aap, 574, A3

\bibitem[{{Peter} {et~al.}(2014){Peter}, {Tian}, {Curdt}, {Schmit}, {Innes},
  {De Pontieu}, {Lemen}, {Title}, {Boerner}, {Hurlburt}, {Tarbell}, {Wuelser},
  {Mart{\'{\i}}nez-Sykora}, {Kleint}, {Golub}, {McKillop}, {Reeves}, {Saar},
  {Testa}, {Kankelborg}, {Jaeggli}, {Carlsson}, \&
  {Hansteen}}]{2014Sci...346C.315P}
{Peter}, H., {Tian}, H., {Curdt}, W., {et~al.} 2014, Science, 346, C315

\bibitem[{{Qiu} {et~al.}(2000){Qiu}, {Ding}, {Wang}, {Denker}, \&
  {Goode}}]{2000ApJ...544L.157Q}
{Qiu}, J., {Ding}, M.~D., {Wang}, H., {Denker}, C., \& {Goode}, P.~R. 2000,
  \apjl, 544, L157

\bibitem[{{Reid} {et~al.}(2016){Reid}, {Mathioudakis}, {Doyle}, {Scullion},
  {Nelson}, {Henriques}, \& {Ray}}]{2016ApJ...823..110R}
{Reid}, A., {Mathioudakis}, M., {Doyle}, J.~G., {et~al.} 2016, \apj, 823, 110

\bibitem[{{Rouppe van der Voort} {et~al.}(2016){Rouppe van der Voort},
  {Rutten}, \& {Vissers}}]{2016A&A...592A.100R}
{Rouppe van der Voort}, L.~H.~M., {Rutten}, R.~J., \& {Vissers}, G.~J.~M. 2016,
  \aap, 592, A100

\bibitem[{{Rutten} {et~al.}(2013){Rutten}, {Vissers}, {Rouppe van der Voort},
  {S{\"u}tterlin}, \& {Vitas}}]{2013JPhCS.440a2007R}
{Rutten}, R.~J., {Vissers}, G.~J.~M., {Rouppe van der Voort}, L.~H.~M.,
  {S{\"u}tterlin}, P., \& {Vitas}, N. 2013, Journal of Physics Conference
  Series, 440, 012007

\bibitem[{{Scharmer} {et~al.}(2003){Scharmer}, {Bjelksjo}, {Korhonen},
  {Lindberg}, \& {Petterson}}]{2003SPIE.4853..341S}
{Scharmer}, G.~B., {Bjelksjo}, K., {Korhonen}, T.~K., {Lindberg}, B., \&
  {Petterson}, B. 2003, in Society of Photo-Optical Instrumentation Engineers
  (SPIE) Conference Series, Vol. 4853, Innovative Telescopes and
  Instrumentation for Solar Astrophysics, ed. S.~L. {Keil} \& S.~V. {Avakyan},
  341--350

\bibitem[{{Scharmer} {et~al.}(2008){Scharmer}, {Narayan}, {Hillberg}, {de la
  Cruz Rodr{\'{\i}}guez}, {L{\"o}fdahl}, {Kiselman}, {S{\"u}tterlin}, {van
  Noort}, \& {Lagg}}]{2008ApJ...689L..69S}
{Scharmer}, G.~B., {Narayan}, G., {Hillberg}, T., {et~al.} 2008, \apjl, 689,
  L69

\bibitem[{{Sukhorukov} \& {Leenaarts}(2016)}]{2016arXiv160605180S}
{Sukhorukov}, A.~V. \& {Leenaarts}, J. 2016, ArXiv e-prints
  [\eprint[arXiv]{1606.05180}]

\bibitem[{{Tian} {et~al.}(2016){Tian}, {Xu}, {He}, \&
  {Madsen}}]{2016ApJ...824...96T}
{Tian}, H., {Xu}, Z., {He}, J., \& {Madsen}, C. 2016, \apj, 824, 96

\bibitem[{{Tortosa-Andreu} \& {Moreno-Insertis}(2009)}]{2009A&A...507..949T}
{Tortosa-Andreu}, A. \& {Moreno-Insertis}, F. 2009, \aap, 507, 949

\bibitem[{{Uitenbroek}(2001)}]{2001ApJ...557..389U}
{Uitenbroek}, H. 2001, \apj, 557, 389

\bibitem[{{van Noort} {et~al.}(2005){van Noort}, {Rouppe van der Voort}, \&
  {L{\"o}fdahl}}]{2005SoPh..228..191V}
{van Noort}, M., {Rouppe van der Voort}, L., \& {L{\"o}fdahl}, M.~G. 2005,
  \solphys, 228, 191

\bibitem[{{Vissers} \& {Rouppe van der Voort}(2012)}]{2012ApJ...750...22V}
{Vissers}, G. \& {Rouppe van der Voort}, L. 2012, \apj, 750, 22

\bibitem[{{Vissers} {et~al.}(2013){Vissers}, {Rouppe van der Voort}, \&
  {Rutten}}]{2013ApJ...774...32V}
{Vissers}, G.~J.~M., {Rouppe van der Voort}, L.~H.~M., \& {Rutten}, R.~J. 2013,
  \apj, 774, 32

\bibitem[{{Vissers} {et~al.}(2015){Vissers}, {Rouppe van der Voort}, {Rutten},
  {Carlsson}, \& {De Pontieu}}]{2015ApJ...812...11V}
{Vissers}, G.~J.~M., {Rouppe van der Voort}, L.~H.~M., {Rutten}, R.~J.,
  {Carlsson}, M., \& {De Pontieu}, B. 2015, \apj, 812, 11

\bibitem[{{Watanabe} {et~al.}(2011){Watanabe}, {Vissers}, {Kitai}, {Rouppe van
  der Voort}, \& {Rutten}}]{2011ApJ...736...71W}
{Watanabe}, H., {Vissers}, G., {Kitai}, R., {Rouppe van der Voort}, L., \&
  {Rutten}, R.~J. 2011, \apj, 736, 71

\bibitem[{{Xu} {et~al.}(2011){Xu}, {Fang}, {Ding}, \&
  {Gao}}]{2011RAA....11..225X}
{Xu}, X.-Y., {Fang}, C., {Ding}, M.-D., \& {Gao}, D.-H. 2011, Research in
  Astronomy and Astrophysics, 11, 225

\end{thebibliography}

%\clearpage
\end{document}